# Magnon and Spin Transition Contribution in Heat Capacity of Ferromagnetic Cr-doped MnTe: Experimental Evidence for a Paramagnetic Spin-Caloritronic Effect

Md Mobarak Hossain Polash,[a,b] Morteza Rasoulianboroujeni,[c] and Daryoosh Vashaee[a,b,*]

[a] Department of Materials Science and Engineering, NC State University, Raleigh, NC 27606, US
[b] Department of Electrical and Computer Engineering, NC State University, Raleigh, NC 27606, US
[c] School of Dentistry, Marquette University, Milwaukee, WI 53233, US

We present experimental evidence for the simultaneous existence of the magnons and spin-state transition contributions to the heat capacity in ferromagnetic (FM) Cr-doped MnTe ($T_c$~280K), where the magnon heat capacity is attributed to the observed magnon-bipolar carrier-drag thermopower. The pristine antiferromagnetic (AFM) MnTe shows only a magnon-induced peak in the heat capacity near the Néel temperature, $T_N$~307K. However, Cr-doped MnTe shows a magnon-contributed heat capacity peak at ~293K with an additional peak in the deep paramagnetic domain near 780K. Temperature-dependent magnetic susceptibility reveals that Cr-doping initially creates low-spin (LS) states $Mn^{2+}$ ions into MnTe near and below $T_N$ due to a higher crystal field induced by Cr ions. Above 400K, LS $Mn^{2+}$ ions start converting into high-spin (HS) $Mn^{2+}$ ions. The LS-to-HS transition of $Mn^{2+}$ leads to an excess entropy and hence excess heat capacity contribution in the system. Temperature-dependent X-ray diffraction (XRD) and magnetic field-dependent susceptibility (M-H) confirmed no presence of any structural changes and magnetic polaron, respectively. Both XRD and M-H ensure that the peak of the heat capacity in the paramagnetic domain is originated solely by the spin-state transition. The heat capacity versus temperature was calculated to explain the contribution of each component, including the ones due to the phonons, magnons, spin-transition, Schottky anomaly, and lattice dilation. With the recent advances in spin-caloritronics extending the spin-based effects from magnetic to paramagnetic materials, the data from the heat capacity can play a crucial role to probe the presence of different phenomena such as paramagnon-carrier-drag and spin-entropy thermopowers.

The interest in magnetic semiconductors for thermoelectric research is overgrowing due to the demonstrations of promising spin-caloritronic effects such as spin-Seebeck,[1] magnon-drag,[2,3,4] paramagnon-drag,[5] and spin-entropy[6] effects. Over the last two decades, the progress in thermoelectric research has been dominated primarily by finding methods for reducing the thermal conductivity of the materials such as increasing the phonon scattering via nano inclusions[7,8,9,10,11,12] and nanostructuring.[13,14] Some other techniques have been developed to enhance the power factor with increasing the thermopower such as carrier filtering[15,16,17,18], carrier pocket engineering[19,20,21], complex structures[22,23,24] creation of resonant energy levels close to the band edges[25], and low dimensional structures.[26,27] The spin-based phenomena are of particular interest as they offer alternate routes to enhance the thermopower, which can be combined with nanostructuring methods to reduce the thermal conductivity simultaneously. The paramagnon-drag and spin-entropy, in particular, extend the spin-caloritronic effects to the paramagnetic phase, which opens a much larger domain for thermoelectric materials search and innovation along with the magnetic semiconductors.[4,5,28] In this regard, the progress in magnetic and paramagnetic thermoelectrics relies on complementary spin-related characterizations, in addition to the standard thermoelectric properties measurements, to better understand and benefit from the spin-based effects.

As thermopower is the amount of entropy carried by the carriers, heat capacity is one of the effective methods to probe the presence of different entropy carriers like lattice vibrations (phonons), charge carriers, spin and orbital degree of freedoms.[29] Typically, heat capacity can have contributions from lattice ($C_v$), electronic ($C_e$), magnon ($C_m$), Schottky anomaly ($C_{sc}$), dilation ($C_d$), and spin transition ($C_{st}$), which are directly related to those degrees of freedoms.[30] A distinct sharp peak appears in heat capacity due to the latent heat associated with the first-order transitions like lattice structure or some magnetic phase changes in the materials. In contrast, the second-order transitions like spin-state transitions and Schottky effects show a broad peak in $C_p$, which is related to the changes in the free energy per molecule.[31] From the contributors of the heat capacity, distinct information on energetics,[32] thermal, electrical, and lattice properties[33] can be obtained.

In this work, we demonstrated the existence of the specific heat contributions from both quantized spin-wave or magnons at magnetic transition temperature and spin-state transition in the paramagnetic domain in Cr-doped MnTe (FIG. 1). Interestingly, Cr-doped MnTe showed magnon-bipolar carrier-drag, and spin-transition influenced thermopower at the corresponding temperatures,[4,28] which can be related to the excess entropy of the system created by spin degree of freedoms. Among spin-based contributions, magnonic contribution to the heat capacity has been widely reported;[3,5,31,33;] however, the spin transition contribution has been demonstrated only in a few materials.[31,32] In addition, both of these contributions in the same material was rarely reported[31] and never reported in any MnTe based systems. MnTe is a widely studied material in spintronics and spin-caloritronics research due to its benevolent spin properties.[5,28,34,35,36] The heat capacity of pristine MnTe is shown in FIG. 1 and compared to that of Cr-doped MnTe ($Mn_{0.95}Cr_{0.05}Te$). While MnTe shows a single magnon heat capacity contribution form the spin system near the Néel temperature, $Mn_{0.95}Cr_{0.05}Te$ shows an additional peak in the paramagnetic phase. In the following, we will discuss the nature of this peak.

*Corresponding author; email: dvashae@ncsu.edu

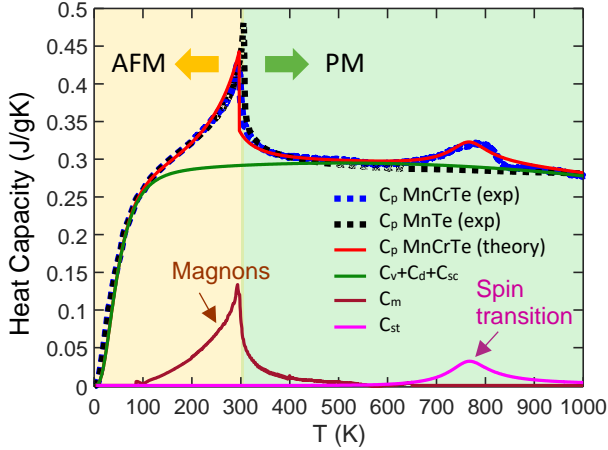

FIG. 1: Heat capacity (Cp) of $Mn_{0.95}Cr_{0.05}Te$ (blue) and MnTe (black) as a function of temperature. Cp plot shows a distinct sharp peak associated with magnon in the magnetic domain (yellow background) and a broad peak associated with the spin-state transition in the paramagnetic domain (green background). Lattice, dilation and Schottky contributions (green) along with magnonic (brown) and spin transition (magenta) contributions, and the total calculated Cp (red) are estimated from mathematical models.[31]

MnTe and 5% Cr-doped MnTe were made from pure (99.99%) elemental powders. Elemental powders were milled for 8hrs in tungsten carbide cup with a proper ratio under the Ar environment to ensure a uniform mixture. The milled powders were sealed in quartz tubes under vacuum and annealed at 850°C for 24hrs in a rocking furnace to achieve uniform doping distribution. The annealed materials were milled again with the same recipe. Finally, the powders were consolidated in graphite dies with spark-plasma-sintering at 950°C with a heating rate of ~60°C/min and soaking time of 20mins under 50MPa pressure and Ar environment. The SPS was performed inside the glovebox and under an inert atmosphere with <0.01ppm $O_2$ and $H_2O$ levels to prevent oxidation of the material. The consolidated ingots were >97% dense. They were characterized by Rigaku Miniflex X-ray diffractometer at room temperature and with PANalytical Empyrean X-ray diffractometer at higher temperatures. Quantum Design DynaCool 12T Physical Property Measurement System (PPMS) was used to measure low-temperature $C_p$ and magnetic susceptibility, and Netzsch differential scanning calorimetry (DSC 404 F1) was used for high-temperature $C_p$ measurement.

The first peak in the heat capacity of both MnTe and $Mn_{0.95}Cr_{0.05}Te$ appeared at ~306K and ~293K, respectively. Both peaks are from the magnon contribution. MnTe is an AFM semiconductor with NiAs structure and a Néel temperature of $T_N$~307K, while $Mn_{0.95}Cr_{0.05}Te$ is an isostructural FM semiconductor with a Curie temperature of $T_C$~280K, as shown in FIG. 2. Cr-doping into MnTe substitutes $Mn^{2+}$ with $Cr^{3+}$ and acts as an electron donor into MnTe.[4,37] It also creates an FM phase by making AFM-FM clustering due to the competition between FM and AFM exchange interactions.[4,38] Alternate potential reasons for the FM phase in Cr-doped MnTe are a canted-spin structure and the spin-polarized hole-mediated ferromagnetic interaction.[4,38] However, the magnon contributed $C_p$ near the Néel temperature of MnTe and the FM trends observed in the inverse magnetic susceptibility (FIG. 2) support the former reason for the magnetic nature of $Mn_{0.95}Cr_{0.05}Te$.[4]

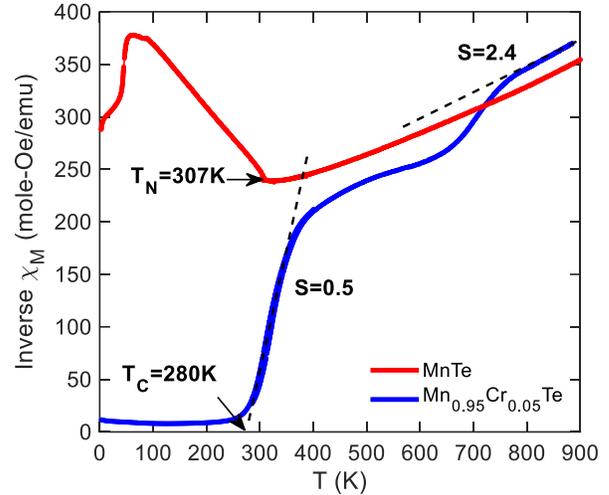

FIG. 2: Inverse magnetic susceptibility of AFM MnTe (red) and FM $Mn_{0.95}Cr_{0.05}Te$ (blue) under the magnetic field of 1000 Oe. The corresponding Néel and Curie temperatures and the spin numbers calculated from the Curie-Weiss law are shown.

Spin numbers for both MnTe and $Mn_{0.95}Cr_{0.05}Te$ are calculated by applying Curie-Weiss law on the inverse susceptibility in the paramagnetic domain. $Mn^{2+}$ in MnTe has a spin number of around 2.4 in the paramagnetic domain, which indicates its high spin (HS) state, while the combination of $Mn^{2+}$ and $Cr^{3+}$ in $Mn_{0.95}Cr_{0.05}Te$ show a distinct spin transition from low spin (LS) of 0.5 to HS of 2.4. Here, all $Cr^{3+}$ ions are always in HS (S=1.5) state due to the $3d^3$ electronic configuration. The lower concentration of $Cr^{3+}$ signifies that $Mn^{2+}$ controls the spin numbers and the corresponding spin-state transition characteristics of $Mn_{0.95}Cr_{0.05}Te$. It is likely that $Cr^{3+}$ ions increase the crystal field in MnTe, which induces the LS state of $Mn^{2+}$ up to about 400K. FIG. 3 illustrates the AFM superexchange interactions between $Mn^{2+}(HS)-Mn^{2+}(HS)$ in MnTe, and $Mn^{2+}(LS)-Mn^{2+}(LS)$ in $Mn_{0.95}Cr_{0.05}Te$ via $Te^{2-}$ p orbitals. The presence of the weak Mn-Mn and Mn-Cr AFM bonding, along with the strong Cr-Cr FM bonding, can reduce the $C_p$ peak in $Mn_{0.95}Cr_{0.05}Te$ compared to that in MnTe.[4]

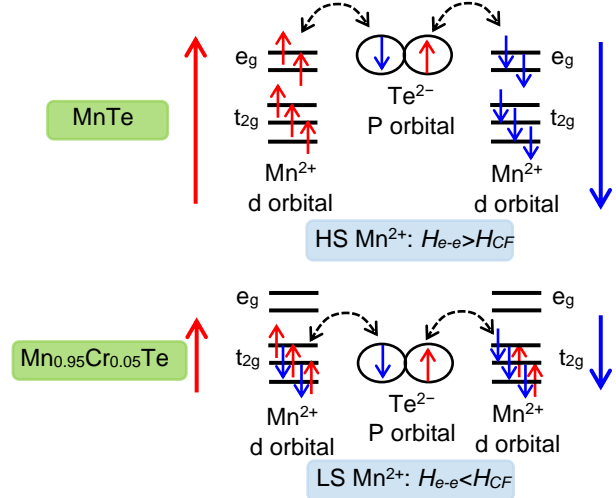

FIG. 3: Mn-Mn AFM superexchange interactions in MnTe and $Mn_{0.95}Cr_{0.05}Te$. Cr-induced high crystal field energy ($H_{CF}$) higher than the electron-electron repulsion energy ($H_{e-e}$) shifts $Mn^{2+}$ from HS into LS state leading to a weak Mn-Mn AFM superexchange interaction.

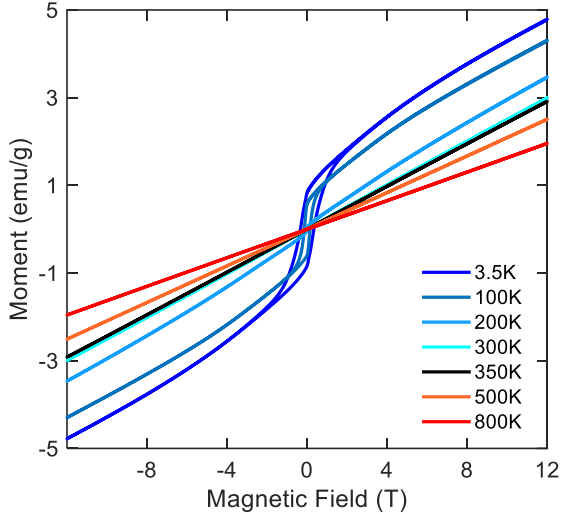

FIG. 4: Magnetic moment of $Mn_{0.95}Cr_{0.05}Te$ vs. magnetic field at different temperatures

At high temperatures, LS $Mn^{2+}$ is transitioned to HS $Mn^{2+}$ by overcoming the crystal field due to both the electron-electron repulsion and the available thermal energy. At comparable crystal field and Hund's exchange energy, both LS and HS ions can coexist, which is often observed in $d^n$ transition-metal compounds ($4 \leq n \leq 7$).[39] With the increase of temperature, three events can occur: a continuous LS to HS transition, a first-order LS-HS transition, or a thermal depletion of HS ground state.[39] However, the broad peak in $C_p$ indicates a continuous LS to HS transition. Other possible reasons for the excess entropy contribution can be the second-order displacive crystal-structure change[31] and magnetic entropy from short-range ferromagnetic correlation in the paramagnetic domain, which can cause a phonon instability and provide some contribution to $C_p$.[31,40] To confirm the underlying reason for the broad $C_p$ peak in $Mn_{0.95}Cr_{0.05}Te$, we carried out both magnetic moment versus magnetic field (M-H) and the XRD at different temperatures. M-H plot can reveal the presence of short-range ferromagnetic correlation, also known as bound magnetic polaron (BMP).[40] However, the M-H trends at different temperatures from different magnetic phase domain in FIG. shows that $Mn_{0.95}Cr_{0.05}Te$ is completely paramagnetic above 300K without any trace of BMPs.

To detect any crystal structure change at the temperature range of interest, a heated XRD was performed at 400K, 600K, and 850K. The obtained data are illustrated in FIG.. The XRD at different temperatures has the same features, which indicates there are no crystal changes. The lack of evidence for BMPs and crystal change suggests that the broad peak of $C_p$ is due to the LS to HS transition of $Mn^{2+}$, which is happening at this range of temperature. Based on the observations from XRD, we can also eliminate the Jahn-

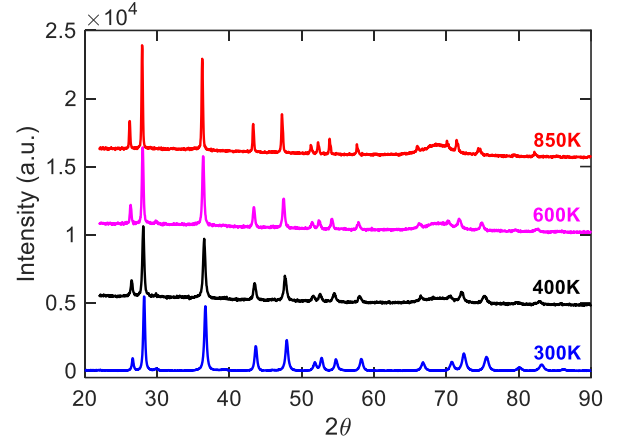

FIG. 5: XRD patterns of $Mn_{0.95}Cr_{0.05}Te$ at different temperatures.

Teller (JT) effect as a contributing factor in $C_p$.

Due to the similarities of $Mn_{0.95}Cr_{0.05}Te$ heat capacity trends and those of MnTe, except a smaller magnon induced peak at ~293K and the broad peak at a higher temperature, lattice, Schottky, and dilation contributions to $C_p$ can be considered to be similar for both materials, and those contributions have been already reported for MnTe.[33] The similar non-spin heat capacity contribution in $Mn_{0.95}Cr_{0.05}Te$ as MnTe, and its small electronic contribution to $C_p$, support that the differences are mainly due to the magnon and spin-state transition contributions. By eliminating the contribution of non-spin-based contributions from $C_p$, the maximum magnon heat capacity ($C_m$) for MnTe is found to be approximately 0.22 J/gK. In comparison, the maximum $C_m$ for $Mn_{0.95}Cr_{0.05}Te$ is estimated to be ~0.16 J/gK. The exchange interaction energies are expected to be less in $Mn_{0.95}Cr_{0.05}Te$ than those in MnTe due to the LS state of $Mn^{2+}$ in $Mn_{0.95}Cr_{0.05}Te$. Therefore, the characteristic temperature, $\theta_M$, related to the magnon ensembles in $Mn_{0.95}Cr_{0.05}Te$ should be less than that in MnTe, because $\theta_M \propto aJS\, k_M/k_B$.[41] Here, $a$ is the distance between the nearest neighbors, $J$ is the exchange interaction energy, $S$ is the spin number, $k_M$ is the maximum value of the wavevector, which can be obtained from either lattice parameters or the carrier concentration, and $k_B$ is the Boltzmann constant.[41] For AFM semiconductors, heat capacity contribution from AFM magnon changes as $C_{mag} \propto T^3$, when $T<<\theta_M$.[41,42]

In Cr-doped MnTe, initially, $Mn^{2+}$ ions are in LS state; hence, there is a distinct exchange interaction between two adjacent LS $Mn^{2+}$ ions, which is denoted as $J_0$ [see Supplementary Material (SM)]. Once the LS $Mn^{2+}$ ions transition to HS state, the exchange interaction energy between two adjacent $Mn^{2+}$ HS ions is different from $J_0$. Exchange interaction energy for nearest-neighbor HS ions is denoted with $J$, which is calculated using the following expression:[31,43]

$$J = \Delta - k_B T L - 3 N k_B T \Delta\theta_D^2 / \theta_D \quad (1)$$

In which $\Delta$ is the energy gap between LS and HS ground state, $N$ is the number of ions per molecule, $\Delta\theta_D$ is the difference of the Debye temperatures ($\theta_D$) between LS and HS states, $Z$ is the partition function, and $T$ is the temperature.[31] $L$ is a function calculated from the partition function of the HS state [SM]. A physical explanation for $\Delta\theta_D$ is discussed in SM. Based on Bragg-Williams approximation, free energy per molecule due to the spin transition is:[43]

$$f = (\Delta - k_B T L)n - J n^2 \\ - k_B T [n \ln n - (1-n)\ln(1-n)] \quad (2)$$

where $n$ is the HS fraction in the material. The calculated $n$ and $f$ are illustrated in FIG 6. From the free energy, heat capacity at constant volume due to the spin transition can be obtained by using $C_{v,st} = -T \left(\frac{\partial^2 f}{\partial T^2}\right)_v$.[31]

Based on the parameters associated with $Mn^{2+}$ ions,[31] along with some fitting parameters, different magnon, and spin transition-related parameters are calculated, which are shown in Table I. It is worth mentioning that both the peak and shape of the spin-transition contributed heat capacity are very sensitive to the fitting parameters, which does not allow us to vary those parameters more than 10-15%. This implies that the fitting parameters cannot be too far from the real physical values. The estimated magnon and spin transition heat capacity contributions along with the total heat capacity (considering non-spin heat capacity contributions of MnTe) are illustrated in FIG 1. A good agreement was found with the experimental data. Details of numerical modeling are discussed in SM.

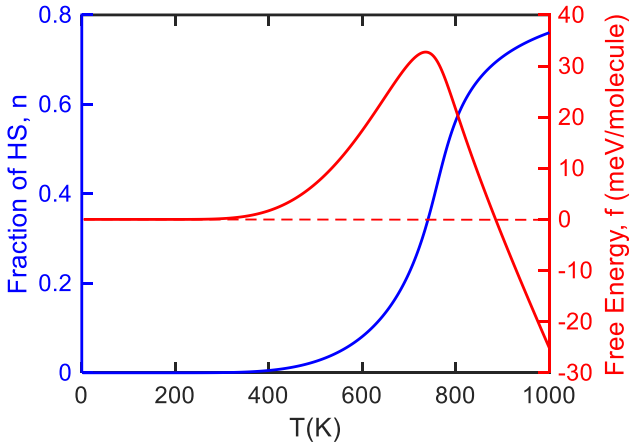

FIG. 4: Fraction of HS ions, $n$ (left axis) and free energy per molecule, $f$ (right axis) in $Mn_{0.95}Cr_{0.05}Te$.

The trends of heat capacity data (FIG. 1), magnetic susceptibility (FIG. 2), and the calculated HS faction (FIG. 6) versus temperature agree that the spin transition happens approximately over the temperature range of 400K - 800K. Nevertheless, the estimated HS fraction extends to a somewhat higher temperature. According to FIG. 2, a significant spin number change occurs at around 400K, while FIG. 6 shows only the onset of LS-to-HS at this temperature, which may arise an ambiguity. It should be noted that we have fitted the heat capacity based on the dominant $Mn^{2+}$ LS-to-HS transition; however, the spin transition in Cr-doped MnTe is more complicated and requires a more detailed theoretical study. The magnetic susceptibility trend of Cr-doped MnTe is associated with several spin-transition phenomena. $Mn_{0.95}Cr_{0.05}Te$ consists of two 3d ions, namely $Cr^{3+}$ and $Mn^{2+}$, which can transition to $Cr^{2+}$ and $Mn^{3+}$ due to the charge transfer reactions. The XRD data show a small trace of CrTe impurity phase, which supports the presence of $Cr^{2+}$.[4,28] The fact that the $C_p$ shows no significant peak associated with the transition near 400K may also be due to the measurement tolerances. While the heat capacity trend is dominated by the majority of $Mn^{2+}$ ions, the other ions can have nontrivial contributions in the more sensitive magnetic susceptibility measurement. This hypothesis is further evidenced by the presence of several distinct peaks for 3d ions in $Mn_{0.95}Cr_{0.05}Te$ reported by high precision spectroscopy measurements.[4,28]

Table I: The estimated magnon and spin-related parameters for MnTe and $Mn_{0.95}Cr_{0.05}Te$ from experimental data and the numerical model. Experimental values are given in parenthesis

| Parameters | MnTe | $Mn_{0.95}Cr_{0.05}Te$ |
|---|---|---|
| $J$ (meV) (LS $Mn^{2+}$) | -- | −16.17 |
| $C_{st}$ (J/mol-K) | 0 | 0.034 (0.032) |
| $f$ (meV) | 0 | 33 |
| $\theta_D$ (K) | 140 | 140 |
| $\Delta\theta_D$ (K) | -- | 7 |
| $\Delta_{LS\text{-}HS}$ (meV) | -- | 255 |
| $\varepsilon$ (meV) | -- | 15 |

In summary, two spin-based heat capacity contributions were demonstrated in Cr-doped MnTe, where the former contribution is coming from magnon and the later from the spin transition in the paramagnetic domain. Both spin-based components create excess spin entropy contributing to spin-caloritronic thermopower enhancement in $Mn_{0.95}Cr_{0.05}Te$. The $C_p$ peak due to magnons is almost identical for both MnTe and Cr-doped MnTe. In $Mn_{0.95}Cr_{0.05}Te$, this peak is attributed to the weak AFM exchange interaction. The paramagnetic $C_p$ peak is associated with $Mn^{2+}$ LS, induced by a higher crystal field in $Mn_{0.95}Cr_{0.05}Te$, to $Mn^{2+}$ HS transition at ~780K. The magnetic susceptibility data agree with the spin transition occurring over the temperature range of 400 K to 800 K. The free energy, heat capacity components, and the faction of HS $Mn^{2+}$ ions were calculated, which quantified the heat capacity contributions due to magnons and the spin transitions.

The details of the heat capacity components calculations are referred to the supplementary material. The data that supports the findings of this study are available within the article and its supplementary material.

This study is partially based upon work supported by the Air

Force Office of Scientific Research (AFOSR) under contract number FA9550-19-1-0363 and the National Science Foundation (NSF) under grant numbers ECCS-1351533, ECCS-1515005, and ECCS-1711253.

**Magnon and Spin Transition Contribution in Heat Capacity of Ferromagnetic Cr-doped MnTe: Experimental Evidence for a Paramagnetic Spin-Caloritronic Effect**

Md Mobarak Hossain Polash,[a,b] Morteza Rasoulianboroujeni,[c] and Daryoosh Vashaee[a,b]

[a]Department of Materials Science and Engineering, North Carolina State University, Raleigh, NC 27606, US
[b]Department of Electrical and Computer Engineering, North Carolina State University, Raleigh, NC 27606, US
[c]School of Dentistry, Marquette University, Milwaukee, WI 53233, US

**S1 Numerical Model to Calculate Spin Transition Contributed Heat Capacity**

For theoretical estimation of the spin-transition contributed heat capacity of $Mn_{0.95}Cr_{0.05}Te$, a mathematical formulation given by H. J. Krokoszinski et al[31] and R. Zimmermann et al[46] was adopted in this work. As mentioned in the paper, low spin (LS) to high spin (HS) transition of a magnetic ion of material causes a change in the free energy per molecule. The difference in free energy provides an excess entropy into the system, which appears as a distinct broad peak in the heat capacity. Therefore, the estimation of the spin transition heat capacity requires the evaluation of the free energy associated with LS to HS transition in LS magnetic ensembles, which can be formulated based on Bragg-Williams approximation given by:[40]

$$f = (\Delta - k_B T L)n - Jn^2 - k_B T[n \ln n - (1-n)\ln(1-n)] \quad (S1)$$

Here, $\Delta$ is the energy gap between LS and HS ground state energy levels, $n$ is the fraction of HS in LS ensembles, $J$ is the temperature-dependent exchange energy, $k_B$ is the Boltzmann constant, and $L$ is a function calculated from the partition function for HS state, $Z$. The fraction of HS, $n$, can be derived from either the total spin number of the system along with the spin numbers of LS and HS ions or a self-consistent relation between $n$ and $Z$.[29] In this work, we calculate $n$ from the self-consistent relationship given as,[29,40]

$$n = \frac{Z}{Z + \exp\left(\frac{\Delta - 2Jn}{k_B T}\right)} \quad (S2)$$

Exchange energy, $J$ in (S2) is determined from the following relation,[31]

$$J = J_0 - c_2 N k_B \frac{\Delta \theta_D^2}{\theta_D} \quad (S3)$$

Here, $J_0$ is the exchange energy between HS ions, which is calculated from the temperature at n=1/2 using the relation:[31]

$$J_0 = \Delta - k_B T_{\frac{1}{2}} L\left(T_{\frac{1}{2}}\right) - c_2\left(T_{\frac{1}{2}}\right) N k_B \frac{\Delta \theta_D^2}{\theta_D} \quad (S4)$$

Where $N$ is the number of ions per molecule, $\theta_D$ is the Debye temperature, and $\Delta\theta_D$ is the difference in the Debye temperature between LS and HS magnetic ions. $L$, $c_1$, and $c_2$ are temperature-dependent functions given by[31]

$$L = \ln Z - c_1 N \frac{\Delta\theta_D}{T} \quad (S5)$$

$$c_1 = \begin{cases} 0.21 - \frac{2.895T}{\theta_D} & T < \frac{\theta_D}{2} \\ 0 & otherwise \end{cases}$$

$$c_2 = \begin{cases} \frac{3T}{\theta_D} & for\ T < \frac{\theta_D}{2} \\ 0 & otherwise \end{cases}$$

$n$ and $T_{\frac{1}{2}}$ are calculated self-consistently, from which $J_0$ and $J$ are estimated. We assumed a similar Debye temperature as MnTe ($\theta_D \sim 140K$) for the Cr-doped MnTe in LS state, as we observed identical phonon heat capacity contribution for both MnTe and Cr-doped MnTe. Therefore, Debye temperature is not a fitting parameter here. $\Delta\theta_D$, however, is a fitting parameter, which we estimated to be ~5% of the Debye temperature. Details on the physical explanation for the difference in the Debye temperature between LS and HS can be found in ref. 40 [section 3, J. Phys. Chm. solids, 38(7), 779-788 (1977)]. In a system having LS-to-HS transition, the modification in the exchange interaction energy can modify the phonon characteristics (not a significant amount). Therefore, the Debye temperature can be modified by a small amount (for Cr-doped, change is around 5%). As the bond strength of the ions is a function of the exchange interaction energy between nearest-neighbor magnetic ions, the spin-state transition can alter the exchange interaction energy and hence the bond strength. This variation in the exchange energy introduces a change in free energy due to the spin-state transition and hence provide excess entropy in the system.

For $Mn_{0.95}Cr_{0.05}Te$, we estimated the partition function, $Z$, based on the following relation,[31,46]

$$Z = \sum_i^{N_{HS}} exp\left(-\frac{E_i - \Delta}{k_B T}\right) \qquad (S6)$$

In which, $N_{HS}$ is the available number of HS energy levels, and $E_i$ is the corresponding energy. The estimation of the available HS energy states depends on the consideration of crystal field, spin-orbit coupling, and zero-field splitting effects. These effects can lift the degeneracy of the energy levels of a magnetic ion and, therefore, can create different $E_i$'s for different energy levels. In $Mn_{0.95}Cr_{0.05}Te$, several 3d magnetic ions can exist, namely, $Mn^{2+}$, $Mn^{3+}$, $Cr^{2+}$, and $Cr^{3+}$. $Cr^{3+}$ ions are coming from the common oxidation state of Cr, while $Mn^{3+}$ and $Cr^{2+}$ ions are due to the charge transfer reaction between $Mn^{2+}$ and $Cr^{3+}$ ($Mn^{2+} + Cr^{3+} \rightarrow Mn^{3+} + Cr^{2+}$). Due to the low amount of $Mn^{3+}$ and $Cr^{2+}$ ions from low Cr doping concentration, and that $Cr^{3+}$ is often at HS state due to its $3d^3$ electronic configuration, $Mn^{2+}$ dominates the overall spin transition of $Mn_{0.95}Cr_{0.05}Te$. Due to the crystal/ligand field, $3d^5$ orbitals of $Mn^{2+}$ ions split into $e_g$ and $t_{2g}$ orbitals, and further lifting of the degeneracy is caused by the spin-orbit coupling (LS) or zero-field splitting (ZFS) interaction. Based on the consideration of crystal symmetry, ligand field, and LS/ZFS, the number of available HS energy levels for $Mn^{2+}$ is found to be 10, as shown in FIG. S1.

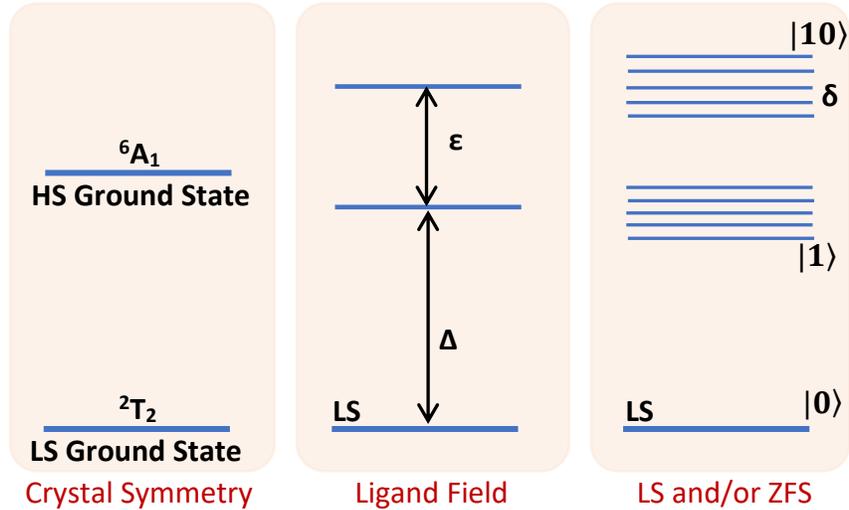

FIG. S1: The available LS and HS energy levels due to the crystal symmetry, ligand field, and spin-orbit coupling (LS) or zero-field splitting (ZFS) interaction.

According to the energy level diagram shown above, we estimated the partition function for HS $Mn^{2+}$ as,

$$Z = \sum_{|1\rangle}^{|5\rangle} exp\left(-\frac{\Delta \pm i\delta - \Delta}{k_B T}\right) + \sum_{|6\rangle}^{|10\rangle} exp\left(-\frac{\Delta + \varepsilon \pm i\delta - \Delta}{k_B T}\right) \qquad (S7)$$

Considering δ is a small number, we can simplify the partition function as:

$$Z = 5 + 5\exp\left(-\frac{\varepsilon}{k_B T}\right) \quad (S8)$$

Based on the above partition function, the fraction, and exchange energy for HS ions were estimated self-consistently. From the results, the free energy of HS ions into LS system was calculated from (S1) and illustrated in FIG. 6. Throughout the calculation, Δ, ε, and $\Delta\theta_D$ were considered as fitting parameters to obtain the broad peak position and the trends of the spin transition heat capacity. For higher spin degeneracy of HS state, $\Delta\theta_D$ can be as low as 5% of $\theta_D$.[31] It is worth mentioning that the peak and shape of the spin-transition contributed heat capacity are very sensitive to the fitting parameters, which does not allow us to vary those parameters more than 10-15%. This implies that the fitting parameters from the modeling cannot be too far from the actual physical values. From the free energy, the heat capacity due to the spin transition can be estimated from:[31]

$$C_{v,st} = -T\left(\frac{\partial^2 f}{\partial T^2}\right)_v \quad (S9)$$

Considering (S1), (S9) can be elaborated by the following expression,[31]

$$C_{v,st} = 2k_B T \frac{dn}{dT}\left[\ln\frac{Z(1-n)}{n} + \frac{\varepsilon}{k_B T}\frac{Z-1}{Z} + 5.79\frac{\Delta\theta_D}{\theta_D} + 12n\left(\frac{\Delta\theta_D}{\theta_D}\right)^2\right] + \left(\frac{dn}{dT}\right)^2\left(2JT - \frac{k_B T^2}{n(1-n)}\right)$$
$$+ k_B T^2 \frac{d^2n}{dT^2}\left(\ln\frac{1-n}{n} + L - \frac{\Delta - 2Jn}{k_B T}\right) + nk_B\left(\frac{\varepsilon}{k_B T}\right)^2\frac{Z-1}{Z^2} \quad (S10)$$

$$C_{v,st} = p1 + p2 + p3 + p4 \quad (S11)$$

The individual *p*'s as a function of temperature are shown in FIG. S2. For $Mn_{0.95}Cr_{0.05}Te$, p1 and p2 dominate over p3 and p4, and spin transition contributed heat capacity follows the trend of p1 (shown in FIG. 1). All the fitting parameters are listed in Table I, along with other estimated spin-based parameters from the experimental data.

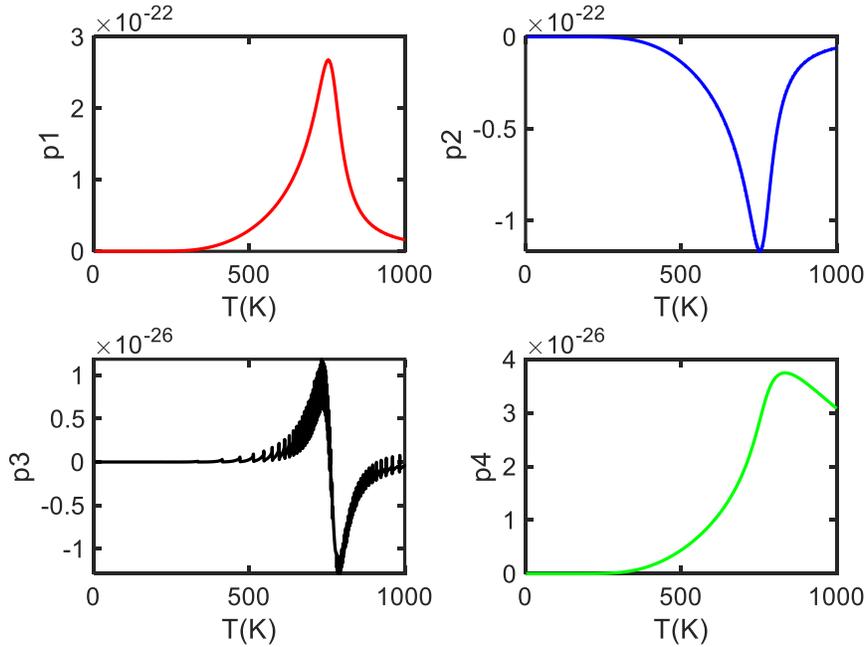

FIG. S2: Different temperature-dependent terms in constant volume heat capacity due to the spin transition given in (10) and (11).

## S2 Estimation of Non-Spin Heat Capacity Contributions

To obtain the spin-based heat capacity contributions of $Mn_{0.95}Cr_{0.05}Te$, non-spin based heat capacity contributions, namely lattice, dilation, and Schottky, must be extracted from the total heat capacity. In a low Cr-doped MnTe like $Mn_{0.95}Cr_{0.05}Te$, crystal structure can be considered identical to the pristine MnTe. Therefore, the same Debye and Einstein temperatures, Gruneisen parameter, and thermal expansivity as MnTe are considered to calculate the lattice and dilation heat capacity components based on the corresponding expressions in ref. 31.

The Schottky heat capacity in $Mn_{0.95}Cr_{0.05}Te$ is supposed to be different than that of MnTe because of the difference in spin state of $Mn^{2+}$ ions. $Mn^{2+}$ is at a low spin state in $Mn_{0.95}Cr_{0.05}Te$ and at a high spin state in MnTe. Based on the expression given in ref. 33, Schottky heat capacity for $Mn_{0.95}Cr_{0.05}Te$ was estimated. Estimated lattice ($C_v$), dilation ($C_d$), and Schottky heat capacity ($C_{Sc}$) along with the calculated spin-based heat capacity contribution ($C_{spin} = C_p - C_v - C_d - C_{Sc}$) are illustrated in FIG. S3.

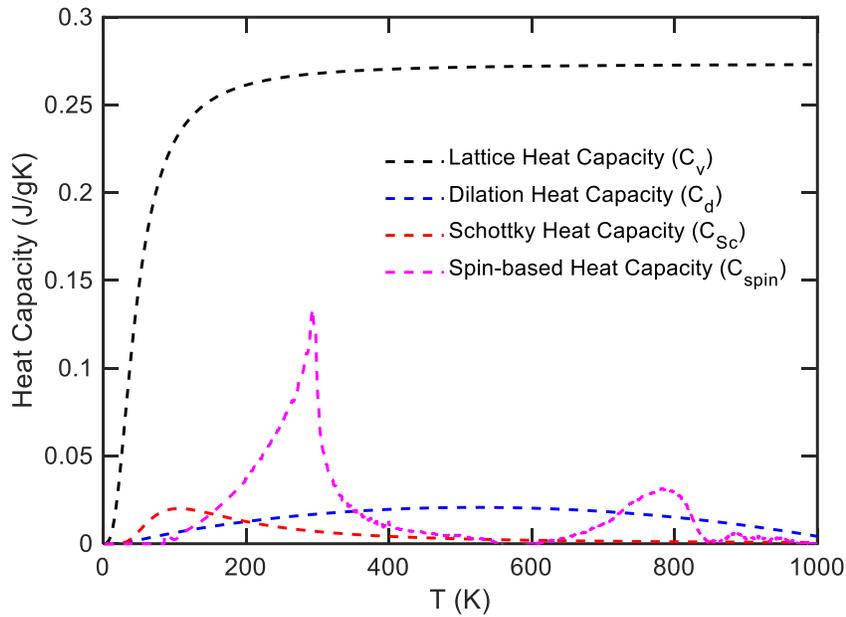

FIG. S3: Different non-spin (lattice, dilation and Schottky) and spin-based (magnon + spin-state transition) heat capacity contributions of $Mn_{0.95}Cr_{0.05}Te$.